\documentclass[twocolumn,twocolappendix]{aastex63}

\received{}
\revised{}
\accepted{}
\shorttitle{H$\beta$ BH Mass and Optical Spectra of $z \sim 6.5$ Quasars}
\shortauthors{Yang et al.}

\def\mgii{\ion{Mg}{2}}
\def\civ{\ion{C}{4}}
\def\cii{\ion{C}{2}}

\def\heii{\ion{He}{2}}

\def\feii{\ion{Fe}{2}}
\def\oiii{\ion{O}{3}}

\begin{document}

\title{A SPectroscopic survey of biased halos In the Reionization Era (ASPIRE): A First Look at the Rest-frame Optical Spectra of $z > 6.5$ Quasars Using JWST}

\correspondingauthor{Jinyi Yang}
\email{jinyiyang@email.arizona.edu}

\author[0000-0001-5287-4242]{Jinyi Yang}
\altaffiliation{Strittmatter Fellow}
\affiliation{Steward Observatory, University of Arizona, 933 N Cherry Avenue, Tucson, AZ 85721, USA}

\author[0000-0002-7633-431X]{Feige Wang}
\affiliation{Steward Observatory, University of Arizona, 933 N Cherry Avenue, Tucson, AZ 85721, USA}

\author[0000-0003-3310-0131]{Xiaohui Fan}
\affiliation{Steward Observatory, University of Arizona, 933 N Cherry Avenue, Tucson, AZ 85721, USA}

\author[0000-0002-7054-4332]{Joseph F. Hennawi}
\affiliation{Department of Physics, University of California, Santa Barbara, CA 93106-9530, USA}
\affiliation{Leiden Observatory, Leiden University, Niels Bohrweg 2, NL-2333 CA Leiden, Netherlands}

\author[0000-0002-3026-0562]{Aaron J. Barth}
\affil{Department of Physics and Astronomy, University of California, Irvine, CA 92697, USA}
\author[0000-0002-2931-7824]{Eduardo Ba\~nados}
\affil{Max Planck Institut f\"ur Astronomie, K\"onigstuhl 17, D-69117, Heidelberg, Germany}
\author[0000-0002-4622-6617]{Fengwu Sun}
\affiliation{Steward Observatory, University of Arizona, 933 N Cherry Avenue, Tucson, AZ 85721, USA}
\author[0000-0003-3762-7344]{Weizhe Liu}
\affiliation{Steward Observatory, University of Arizona, 933 N Cherry Avenue, Tucson, AZ 85721, USA}
\author[0000-0001-8467-6478]{Zheng Cai}
\affiliation{Department of Astronomy, Tsinghua University, Beijing 100084, China}
\author[0000-0003-4176-6486]{Linhua Jiang}
\affiliation{Department of Astronomy, School of Physics, Peking University, Beijing 100871, China}
\affiliation{Kavli Institute for Astronomy and Astrophysics, Peking University, Beijing 100871, China}
\author[0000-0001-5951-459X]{Zihao Li}
\affiliation{Department of Astronomy, Tsinghua University, Beijing 100084, China}
\author[0000-0003-2984-6803]{Masafusa Onoue}
\affiliation{Kavli Institute for Astronomy and Astrophysics, Peking University, Beijing 100871, China}
\affiliation{Kavli Institute for the Physics and Mathematics of the Universe (Kavli IPMU, WPI), The University of Tokyo, Chiba 277-8583, Japan}

\author[0000-0002-4544-8242]{Jan--Torge Schindler}
\affiliation{Leiden Observatory, Leiden University, Niels Bohrweg 2, NL-2333 CA Leiden, Netherlands}
\author[0000-0003-1659-7035]{Yue Shen}
\affiliation{Department of Astronomy, University of Illinois at Urbana-Champaign, Urbana, IL 61801, USA}
\affiliation{National Center for Supercomputing Applications, University of Illinois at Urbana-Champaign, Urbana, IL 61801, USA}
\author[0000-0003-0111-8249]{Yunjing Wu}
\affiliation{Department of Astronomy, Tsinghua University, Beijing 100084, China}
\affiliation{Steward Observatory, University of Arizona, 933 N Cherry Avenue, Tucson, AZ 85721, USA}

\author[0000-0002-7080-2864]{Aklant K. Bhowmick}
\affiliation{Department of Physics, University of Florida, Gainesville, FL, 32611, USA}

\author{Rebekka Bieri}
\affil{Center for Space and Habitability, University of Bern, Gesellschaftsstrasse 6 (G6), Bern, Switzerland}

\author[0000-0002-2183-1087]{Laura Blecha}
\affiliation{Department of Physics, University of Florida, Gainesville, FL 32611-8440, USA}

\author[0000-0001-8582-7012]{Sarah Bosman}
\affil{Max Planck Institut f\"ur Astronomie, K\"onigstuhl 17, D-69117, Heidelberg, Germany}

\author[0000-0002-6184-9097]{Jaclyn B. Champagne}
\affiliation{Steward Observatory, University of Arizona, 933 N Cherry Avenue, Tucson, AZ 85721, USA}

\author{Luis Colina}
\affil{Centro de Astrobiología (CAB), CSIC-INTA, Ctra. de Ajalvir km 4, Torrejón de Ardoz, E-28850, Madrid, Spain}
\affil{International Associate, Cosmic Dawn Center (DAWN)}

\author[0000-0002-7898-7664]{Thomas Connor}
\affiliation{Center for Astrophysics $\vert$\ Harvard\ \&\ Smithsonian, 60 Garden St., Cambridge, MA 02138, USA}
\affiliation{Jet Propulsion Laboratory, California Institute of Technology, 4800 Oak Grove Drive, Pasadena, CA 91109, USA}

\author[0000-0002-6748-2900]{Tiago Costa}
\affiliation{Max-Planck-Institut f\"ur Astrophysik, Karl-Schwarzschild-Stra{\ss}e 1, D-85748 Garching b. M\"unchen, Germany}

\author[0000-0003-0821-3644]{Frederick B.\ Davies}
\affil{Max Planck Institut f\"ur Astronomie, K\"onigstuhl 17, D-69117, Heidelberg, Germany}

\author[0000-0002-2662-8803]{Roberto Decarli}
\affil{INAF--Osservatorio di Astrofisica e Scienza dello Spazio, via Gobetti 93/3, I-40129, Bologna, Italy}

\author[0000-0003-3242-7052]{Gisella De Rosa}
\affiliation{Space Telescope Science Institute, 3700 San Martin Dr, Baltimore, MD 21210}

\author[0000-0002-0174-3362]{Alyssa B.\ Drake}
\affiliation{Centre for Astrophysics Research, Department of Physics, Astronomy and Mathematics, University of Hertfordshire, Hatfield AL10 9AB, UK}

\author[0000-0003-1344-9475]{Eiichi Egami}
\affiliation{Steward Observatory, University of Arizona, 933 N Cherry Avenue, Tucson, AZ 85721, USA}

\author[0000-0003-2895-6218]{Anna--Christina~Eilers}\thanks{Pappalardo Fellow}
\affiliation{MIT Kavli Institute for Astrophysics and Space Research, 77 Massachusetts Ave., Cambridge, MA 02139, USA}

\author[0000-0003-0850-7749]{Analis E. Evans}
\affiliation{Department of Physics, University of Florida, Gainesville, FL, 32611, USA}

\author[0000-0002-6822-2254]{Emanuele Paolo Farina}
\affiliation{Gemini Observatory, NSF's NOIRLab, 670 N A'ohoku Place, Hilo, Hawai'i 96720, USA}

\author{Melanie Habouzit}
\affiliation{Zentrum f\"ur Astronomie der Universit\"at Heidelberg, ITA, Albert-Ueberle-Str. 2, D-69120 Heidelberg, Germany}
\affil{Max Planck Institut f\"ur Astronomie, K\"onigstuhl 17, D-69117, Heidelberg, Germany}

\author[0000-0003-3633-5403]{Zoltan Haiman}
\affil{Department of Astronomy, Columbia University, New York, NY 10027, USA}
\affil{Department of Physics, Columbia University, New York, NY 10027, USA}

\author[0000-0002-5768-738X]{Xiangyu Jin}
\affiliation{Steward Observatory, University of Arizona, 933 N Cherry Avenue, Tucson, AZ 85721, USA}

\author[0000-0003-1470-5901]{Hyunsung D. Jun}
\affil{SNU Astronomy Research Center, Seoul National University, 1 Gwanak-ro, Gwanak-gu, Seoul 08826, Republic of Korea}

\author[0000-0001-6874-1321]{Koki Kakiichi}
\affiliation{Department of Physics, Broida Hall, University of California, Santa Barbara, CA 93106-9530, USA}

\author[0000-0002-7220-397X]{Yana Khusanova}
\affil{Max Planck Institut f\"ur Astronomie, K\"onigstuhl 17, D-69117, Heidelberg, Germany}

\author[0000-0001-5829-4716]{Girish Kulkarni}
\affiliation{Tata Institute of Fundamental Research, Homi Bhabha Road, Mumbai 400005, India}

\author{Federica Loiacono}
\affil{INAF--Osservatorio di Astrofisica e Scienza dello Spazio, via Gobetti 93/3, I-40129, Bologna, Italy}

\author{Alessandro Lupi}
\affil{Dipartimento di Fisica ``G. Occhialini'', Universit\`a degli Studi di Milano-Bicocca, Piazza della Scienza 3, I-20126 Milano, Italy}

\author[0000-0002-5941-5214]{Chiara Mazzucchelli}
\affiliation{Instituto de Estudios Astrof\'{\i}sicos, Facultad de Ingenier\'{\i}a y Ciencias, Universidad Diego Portales, Avenida Ejercito Libertador 441, Santiago, Chile}

\author[0000-0003-0230-6436]{Zhiwei Pan}
\affiliation{Department of Astronomy, School of Physics, Peking University, Beijing 100871, China}
\affiliation{Kavli Institute for Astronomy and Astrophysics, Peking University, Beijing 100871, China}

\author[0000-0003-2349-9310]{Sof\'ia Rojas-Ruiz}
\altaffiliation{Fellow of the International Max Planck Research School for Astronomy and Cosmic Physics at the University of Heidelberg (IMPRS--HD)}
\affil{Max Planck Institut f\"ur Astronomie, K\"onigstuhl 17, D-69117, Heidelberg, Germany}

\author[0000-0002-0106-7755]{Michael A. Strauss}
\affiliation{Department of Astrophysical Sciences, Princeton University, Princeton, NJ 08544 USA}

\author[0000-0003-0747-1780]{Wei Leong Tee}
\affiliation{Steward Observatory, University of Arizona, 933 N Cherry Avenue, Tucson, AZ 85721, USA}

\author[0000-0002-3683-7297]{Benny Trakhtenbrot}
\affiliation{School of Physics and Astronomy, Tel Aviv University, Tel Aviv 69978, Israel}

\author[0000-0002-6849-5375]{Maxime Trebitsch}
\affil{Kapteyn Astronomical Institute, University of Groningen, P.O. Box 800, 9700 AV Groningen, The Netherlands}

\author[0000-0001-9024-8322]{Bram Venemans}
\affiliation{Leiden Observatory, Leiden University, Niels Bohrweg 2, NL-2333 CA Leiden, Netherlands}

\author[0000-0001-9191-9837]{Marianne Vestergaard}
\affiliation{The Niels Bohr Institute, University of Copenhagen, Denmark}
\affiliation{Steward Observatory, University of Arizona, 933 N Cherry Avenue, Tucson, AZ 85721, USA}

\author{Marta Volonteri}
\affil{Institut d'Astrophysique de Paris, Sorbonne Universit\'e, CNRS, UMR 7095, 98 bis bd Arago, 75014 Paris, France}

\author[0000-0003-4793-7880]{Fabian Walter}
\affil{Max Planck Institut f\"ur Astronomie, K\"onigstuhl 17, D-69117, Heidelberg, Germany}

\author[0000-0002-0125-6679]{Zhang-Liang Xie}
\affil{Max Planck Institut f\"ur Astronomie, K\"onigstuhl 17, D-69117, Heidelberg, Germany}

\author[0000-0002-5367-8021]{Minghao Yue}
\affiliation{MIT Kavli Institute for Astrophysics and Space Research, 77 Massachusetts Ave., Cambridge, MA 02139, USA}
\affiliation{Steward Observatory, University of Arizona, 933 N Cherry Avenue, Tucson, AZ 85721, USA}

\author[0000-0002-4321-3538]{Haowen Zhang}
\affiliation{Steward Observatory, University of Arizona, 933 N Cherry Avenue, Tucson, AZ 85721, USA}

\author[0000-0002-0123-9246]{Huanian Zhang} 
\affil{Department of Astronomy, Huazhong University of Science and Technology, Wuhan, 430074, China}

\author[0000-0002-3983-6484]{Siwei Zou}
\affiliation{Department of Astronomy, Tsinghua University, Beijing 100084, China}



\begin{abstract}
Studies of rest--frame optical emission in quasars at $z>6$ have historically been limited by the wavelengths accessible by ground-based telescopes. The James Webb Space Telescope (JWST) now offers the opportunity to probe this emission deep into the reionization epoch. We report the observations of eight quasars at $z>6.5$ using the JWST/NIRCam Wide Field Slitless Spectroscopy, as a part of the ``A SPectroscopic survey of biased halos In the Reionization Era (ASPIRE)" program. Our JWST spectra cover the quasars' emission between rest frame $\sim$ 4100 and 5100 \AA. The profiles of these quasars' broad H$\beta$ emission lines span a FWHM from 3000 to 6000 $\rm{km~s^{-1}}$. The H$\beta$-based virial black hole (BH) masses, ranging from 0.6 to 2.1 billion solar masses, are generally consistent with their \mgii-based BH masses. The new measurements based on the more reliable H$\beta$ tracer thus confirm the existence of billion solar-mass BHs in the reionization epoch. In the observed [\oiii] $\lambda\lambda$4960,5008 doublets of these luminous quasars, broad components are more common than narrow core components ($\le~1200~\rm{km~s^{-1}}$), and only one quasar shows stronger narrow components than broad. Two quasars exhibit significantly broad and blueshifted [\oiii] emission, thought to trace galactic--scale outflows, with median velocities of $-610~\rm{km~s^{-1}}$ and $-1430~\rm{km~s^{-1}}$ relative to the [\cii] $158\,\mu$m line. All eight quasars show strong optical \feii\ emission, and follow the Eigenvector 1 relations defined by low--redshift quasars. The entire ASPIRE program will eventually cover 25 quasars and provide a statistical sample for the studies of the BHs and quasar spectral properties.
\end{abstract}



\section{Introduction} 
Observations of luminous $z \gtrsim 6.5$ quasars have revealed the existence of supermassive black holes (SMBHs) and their massive host galaxies in the reionization epoch. The measurements of high-redshift quasar BH masses suggest that 10$^{8-9}$ solar-mass SMBHs existed as early as 700 million years after the Big Bang, challenging models of early SMBH growth and BH seed formation \citep[e.g.,][]{banados18,yang20,wang21,yang21,farina22}. Measuring BH masses accurately is the key step required to characterize the growth and evolution of early SMBHs. The way to obtain robust BH mass measurements is the reverberation mapping (RM). However, this is not practically possible for $z > 4$ quasars. At higher redshift, the BH mass estimates are based on the so-called single-epoch virial BH mass using quasar broad emission lines.

The observations of quasar spectra in the rest-frame optical have been mostly confined to quasars at $z \lesssim 4$ \citep[e.g.,][]{shen16,matthews21}, because of the difficulty in observing faint objects from the ground at wavelengths longer than the near-infrared K-band.  
Thus, the measurements of SMBHs in quasars at $z>4$ are mainly based on broad emission lines in the rest-frame UV (e.g., \civ\ $\lambda1549$ and \mgii\ $\lambda2800$). 
The \mgii-based BH mass estimators are widely used for the measurements of reionization-era SMBHs \citep[e.g.,][]{shen19, yang21,farina22}. 
However, the study of quasar BH masses at low redshift via RM and the comparison between multiple single-epoch BH mass estimators have established the H$\beta$ line as the most reliable tracer among quasar UV/optical emission lines \citep[e.g.,][]{shen13, wang20}. 
The existing \mgii-based scaling relation is indirectly calibrated based on the H$\beta$ relation derived from RM, which introduces additional uncertainties. 
Therefore, high sensitivity infrared (IR) spectroscopic observations covering the rest-frame optical emission is needed to better estimate the H$\beta$ based BH masses of quasars at the highest redshift. 

ASPIRE (A SPectroscopic survey of biased halos In the Reionization Era) is a JWST Cycle 1 GO program that will observe a sample of 25 $z>6.5$ quasars with JWST NIRCam imaging and Wide Field Slitless Spectroscopy (WFSS) \citep{wang22}. The WFSS observations will provide the spectra in the H$\beta$ region for reionization-era quasars for the first time. In addition, the high quality JWST spectra will allow us to study the spectral properties (e.g., continuum emission, [\oiii] $\lambda\lambda$ 4960, 5008, and optical \feii\ emission) of quasars in the early Universe.
In this paper, we report JWST observations of eight ASPIRE quasars.  
They form the first sample of reionization-era quasars with spectral coverage in the rest-frame optical. We describe the observations and data reduction in Section 2. In Section 3, we report the main results from this early data, including the spectral fitting procedure, the H$\beta$-based BH masses, and the quasar optical emission line properties.  
In this section, we also present detailed analysis of  the blueshifted [\oiii] components and outflows observed in these quasars. We summarize our work in Section 4. All results assume a flat  $\Lambda$CDM cosmology with parameters $\Omega_{\Lambda}$ = 0.7, $\Omega_{m}$ = 0.3, and $h$ = 0.7. 

\section{Observations and Data Reduction}
The eight quasar spectra used in this work are part of the medium size JWST Cycle-1 program ASPIRE \citep[PID 2078, PI: F. Wang;][]{wang22}, which carries out a spectroscopic survey in 25 reionization-era quasar fields using JWST/NIRCam WFSS using the F356W filter, as well as imaging observations in the F115W, F200W, and F356W bands at the same time. This program allows us to statistically investigate both reionization-era quasars and high-redshift galaxies in the quasar fields. The NIRCam spectral and imaging dataset probes the rest-frame optical emission in a large sample of $z> 6.5$ quasars and host galaxies for the first time, which will improve our understanding of early SMBH activity and their host galaxies. The spectroscopic and imaging observations of the quasar fields allow studies of the environments of reionization-era quasars in a statistical manner, which will address the long-standing question of whether luminous quasars reside in the most massive dark matter halos and inhabit large-scale galaxy overdensities in the early Universe.

The eight quasars reported in this work were observed with JWST/NIRCam in 2022 August and September. These targets were observed using the R-grism with F356W filter in long wavelength (LW) and F200W imaging in short wavelength (SW), simultaneously. We performed a 3-point {\tt INTRAMODULEX} primary dither with a {\tt 2-POINT-LARGE-WITH-NIRISS} sub-pixel dither. A {\tt SHALLOW4} readout mode was used with 9 groups per integration and 1 integration per exposure, which resulted in a total exposure time of 2834s. After the grism observation, direct image and out-of-field images were taken in F356W (LW) and F115W (SW) with a total exposure time of 1417s. 
For the analysis in this work, we only use the WFSS F356W spectral data of the quasar targets. Analysis of the spectra of galaxies in the quasar fields and broad-band imaging data will be presented in a set of subsequent papers. 

The data reduction of imaging and WFSS data will be presented by \cite{wang22} in detail and we will briefly describe related steps below. 
The reduction of WFSS data used in this work was based on the JWST pipeline 1.7.2 and calibration reference data `jwst\_0989.pmap'. We started with {\tt calwebb\_detector1} to process the raw data and applied a custom step to {\tt *\_rate.fits} to remove the $1/f$ noise feature \citep{schlawin20}. 
We performed flat-fielding and assigned WCS for exposures, and then generated master background models based on all existing ASPIRE observations. We constructed a tracing model and a dispersion model following the method used in \cite{sun22}, and applied 2D spectral extraction to background subtracted data. We used updated sensitivity functions based on Cycle 1 calibration programs (PID 1536, 1537, and 1538). Then the 1D spectra were extracted from individual exposures using optimal extraction \citep{horne86} and coadded with inverse variance weighting. The 1D spectra of the eight ASPIRE quasars are shown in Figure \ref{fig:spec01}.

\section{Rest-frame Optical Spectra of Reionization-era Quasars}
In this section, we describe our spectral fitting procedure and the main results derived from this early sample. We measure the BH masses based on the H$\beta$ line width  and continuum luminosity, and compare them with the measurements derived from their rest-frame UV spectra. 
We also discuss the profile of the [\oiii] emission lines and search for potential galactic-scale outflows. We note that the data reduction procedure of this new instrument is still in development, and the spectra might be slightly affected by future changes of the reduction pipeline but the main findings in this paper will not be significantly affected. The final JWST ASPIRE dataset will be presented after this program has been completed.

\subsection{Spectra Fitting}
\begin{figure*}
\centering 
\epsscale{1.19}
\plotone{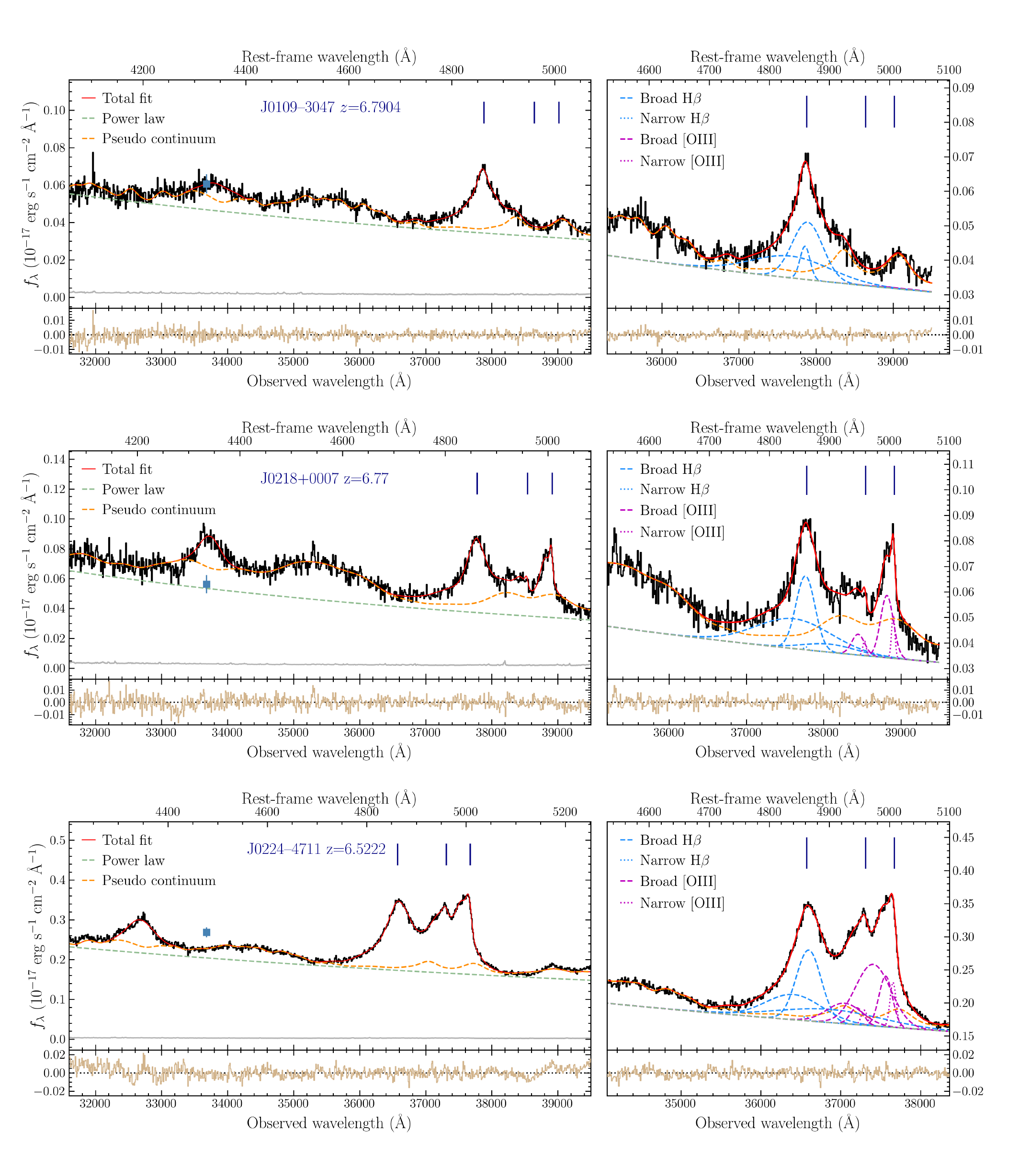} 
\caption{The JWST NIRCam WFSS F356W spectra of eight ASPIRE quasars (black line) with spectral uncertainty (grey), ordered by RA. The red solid lines  denote the best total fits. The best-fits of different spectral components are shown with dashed and dotted lines. For each object, the left panel presents the entire spectrum with total fit, power-law continuum, and pseudo continuum (i.e., power-law plus \feii\ emission). The right panel shows the zoomed-in region of H$\beta$ and [\oiii] lines. The systemic redshifts of the quasars are based on their [\cii] line measurements from ALMA observations. The blue squares with error bars are the photometric data in the WISE $W1$ 3.4 $\mu$m band. The panels below the spectra present the residuals (data--model) of each best-fit model.}
\label{fig:spec01}
\end{figure*}

\addtocounter{figure}{-1}
\begin{figure*}
\centering 
\epsscale{1.19}
\plotone{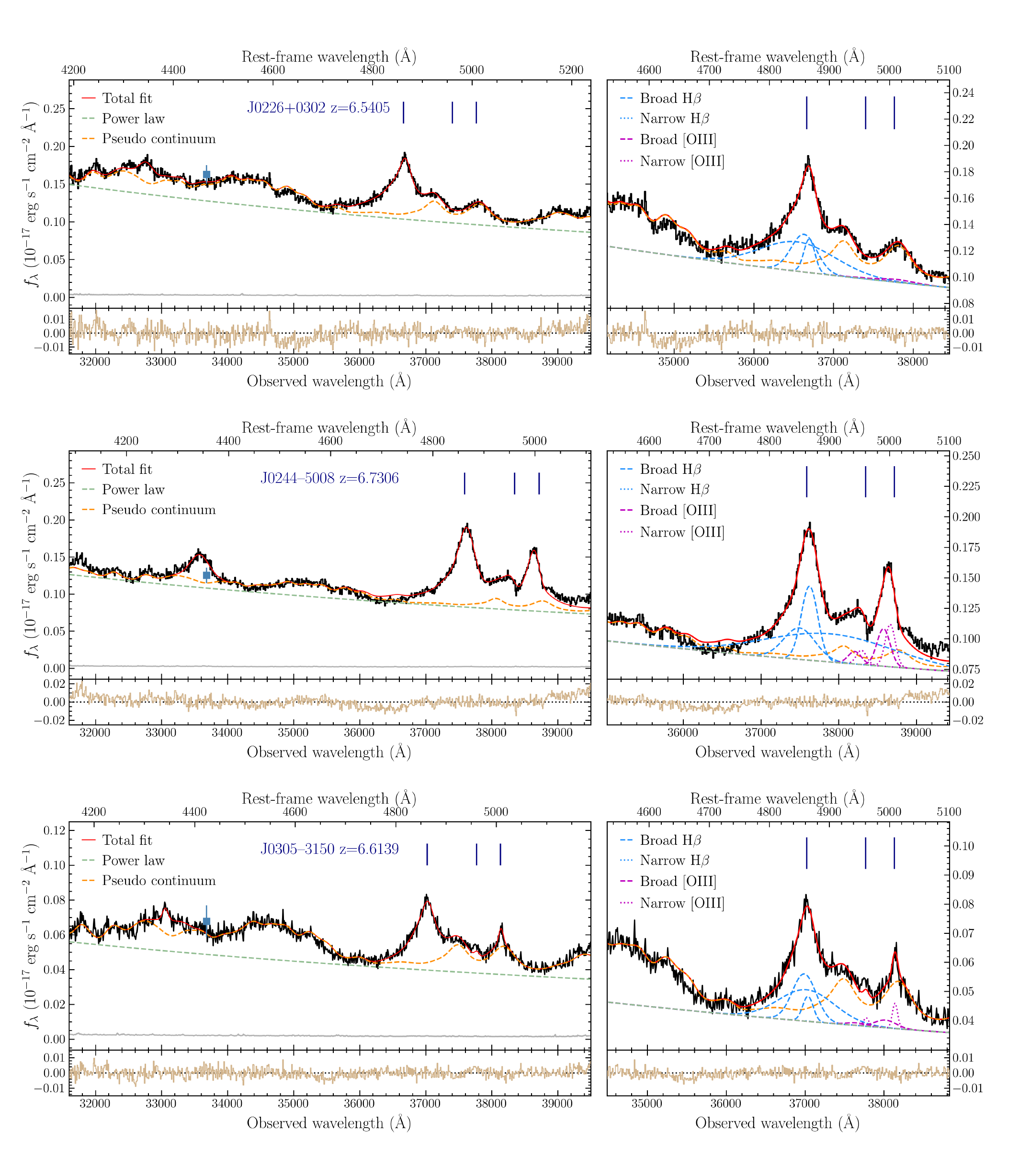} 
\caption{Continued}
\label{fig:spec01}
\end{figure*}

\addtocounter{figure}{-1}
\begin{figure*}
\centering 
\epsscale{1.19}
\plotone{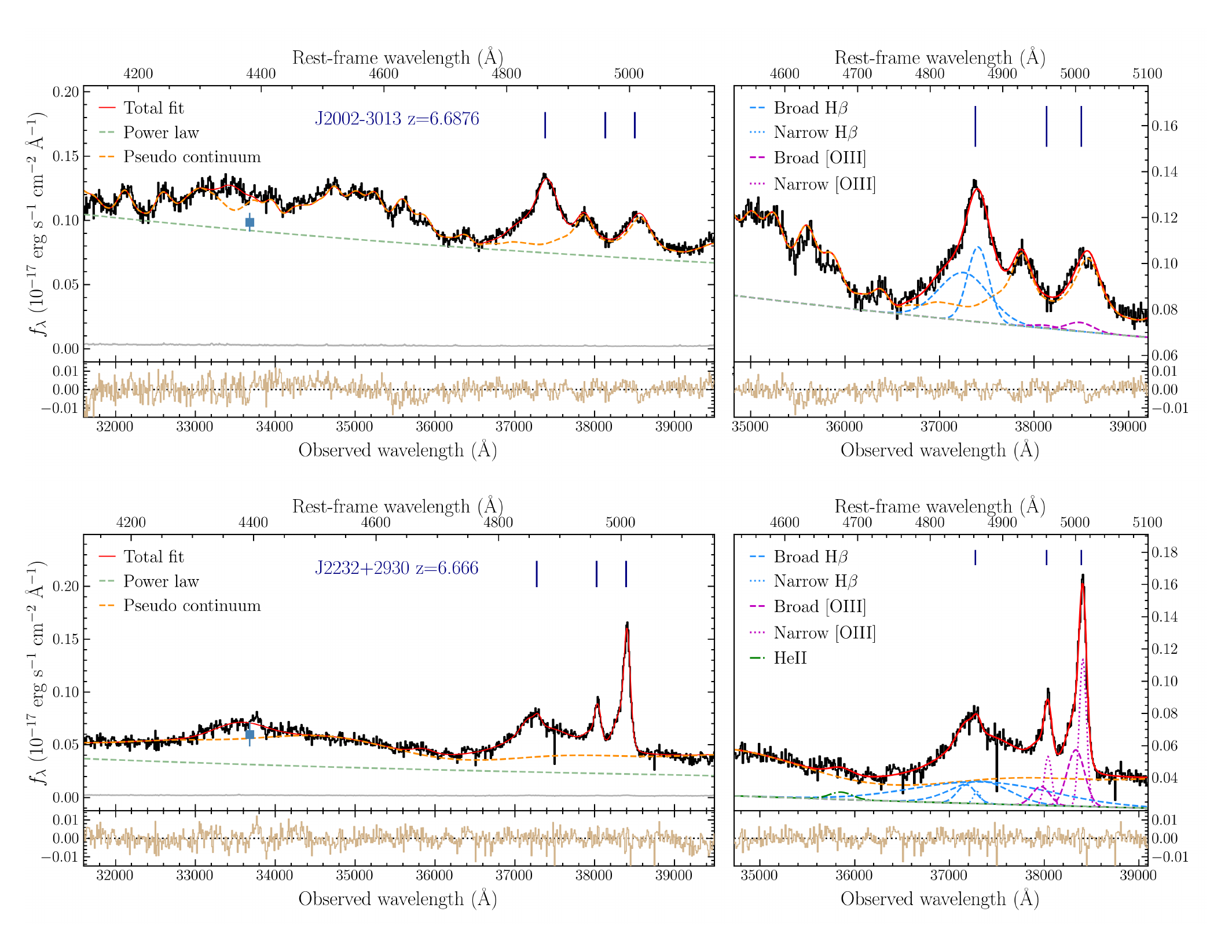} 
\caption{Continued}
\label{fig:spec01}
\end{figure*}

We perform spectral fitting for each WFSS spectrum using a model consisting of continuum and emission lines. The redshift derived from the [\cii] 158 $\mu$m line \citep[][Wang et al. in prep]{venemans20,yang21} is used as the initial redshift.
We use spectral data between observed wavelengths 31600 \AA\ and 39500 \AA, excluding edge regions where the calibration is less certain. This covers the rest-frame wavelength range of $\sim$ 4100 - 5200 \AA\ for these quasars. We use a pseudo-continuum model including a power-law continuum and an empirical optical \feii\ template from \cite{BG92}. To fit this pseudo-continuum, we choose continuum windows that are free of strong emission lines in quasar composite, [4150, 4230], [4435, 4700], and [5100, 5200] \AA\ in the rest frame.
For the two highest-redshift quasars, J0109--3047 and J0218+0007, we adjust the windows due to their limited spectral coverage on the red side. For the former, we only use the first two windows. For the latter, we extend the blue side cutoff of the last window to 5060 \AA.

{\em H$\beta$ and [\oiii] lines:}
We adopt the rest-frame vacuum wavelengths of 4862.68 \AA\ for the H$\beta$ line and 4960.30 \AA\ and 5008.24 \AA\ for the [\oiii] doublet \citep{vandenberk01}.
We employ a line model consisting of one narrow Gaussian component and three broad Gaussian components for the H$\beta$ line \citep[e.g.,][]{shen16}.
For each [\oiii] line, we use one Gaussian for the narrow component (the `core' component) plus a second Gaussian profile for the potentially broadened and blueshifted component. For the object with strong blueshifted [\oiii] wings (i.e., J0224--4711), we include one more Gaussian for its broad [\oiii] component. 
We set a lower limit to the line full width at half maximum (FWHM) of each component to be $200$ km s$^{-1}$. This is based on the instrument resolution ($ R\sim 1600$ at 4 $\mu$m). We also set an upper limit of 1200 km s$^{-1}$ for all narrow components \citep[e.g., ][]{shen11}.

We tie the velocity shift and dispersion of the narrow H$\beta$ to that of the [\oiii] `core' component. 
We assume a fixed line ratio of 3.0 between [\oiii] $\lambda$5008 \AA\ and [\oiii]  $\lambda$4960 \AA\ for both `core' and broad components \citep[e.g.,][]{agnbook}. The velocity shifts of the broad  [\oiii] components of each transition are also tied. 
We do not fix the line ratio between narrow H$\beta$ component and narrow [\oiii] lines, but limit the ratio of the `core' [\oiii] 5008 to the narrow H$\beta$ to $\ge$1. If for an object, there is no narrow [\oiii] 5008 line, we consider that there should also be no narrow H$\beta$. 
For objects with narrow [\oiii] components, we derive [\oiii]-based redshifts using their narrow components, and find that the [\oiii]-based redshifts show good agreement with their redshifts derived from [\cii] 158 $mu$m lines.

{\em \heii\ line:} We test the possibility of including a \heii\ line with a vacuum wavelength of 4687.02 \AA\ by running alternate spectral decomposition. We include one Gaussian profile without constraining the line width and use a continuum window of [4435, 4635] \AA\ instead of [4435, 4700] \AA\ to avoid potential contamination from \heii\ emission. We run the same spectral fitting procedure and fit the \heii\ line together with the H$\beta$ and [\oiii] lines. We find that only one quasar (J2232+2930) has an obvious \heii\ line. Another quasar (J0218+0007) might have a narrow \heii\ line but it only has a significance of $\sim 2\sigma$. For these two, including \heii\ reduces their BH masses by 4\% and 2\%, respectively. For other objects, their best-fit models do not show any need for a \heii\ line, while the shorter continuum windows affect fits of their continuum, especially for $z > 6.7$ quasars. Thus, we finally only apply the \heii\ fit for quasar J2232+2930, and for other quasars we keep the continuum window cutoff of 4700 \AA. Even if there could be some weak \heii\ emission in these quasars, their BH masses would not be affected significantly. 

{\em H$\gamma$ line:}
We also apply a two-Gaussian fit to the H$\gamma$ line at 4341.68 \AA. The two Gaussian profiles are set for one narrow and one broad component. The same limits of narrow line width as used for $H\beta$ are applied. The results are only used for plotting the best-fit but not any science analysis in this work.

We utilize a Monte Carlo (MC) approach to estimate the uncertainties of all spectral measurements, following \cite{yang21} \citep[also][]{shen19, wang20}.
For each spectrum, we generate 50 mock spectra by randomly adding Gaussian noise at each pixel with a standard deviation equal to the spectral error at that pixel. We then perform the same spectral fitting procedure to each mock spectrum. The uncertainty of each spectral measurement is estimated by averaging the 16\% and 84\% percentile deviations from the median value. The best-fit results of all quasars are shown in Figure \ref{fig:spec01}.

\subsection{H$\beta$-based Black Hole Masses}
The single-epoch virial method is widely used for estimating the BH mass of high-redshift quasars, assuming virial motion of the line-emitting gas in the quasar broad line region (BLR) and applying the empirical correlation between the BLR size and quasar continuum luminosity (i.e., the $R-L$ relation). With this method, we can estimate the quasar central BH mass utilizing quasar broad emission lines and continuum luminosity. Quasar UV and optical emission lines, \civ\ $\lambda$1549 \AA, \mgii\ $\lambda$2800 \AA, H$\alpha$ $\lambda$6563 \AA, and H$\beta$ have all been used as virial BH mass tracers \cite[e.g.,][]{mclure02, mclure04, vestergaard02, onken04, vp06, greene05, vo09, shen11}. Calibration coefficients used in these BH mass estimators are determined using low-redshift quasar samples that have BH mass measurements based on RM. Among these lines, H$\beta$ is often thought to be the most reliable tracer from both RM-based measurements and comparison of single-epoch mass estimators \citep[e.g.,][]{shen12}. 

\begin{deluxetable*}{l l l l l l l l l}
\tablecaption{Spectral Fitting and Quasar Properties}
\tablewidth{0pt}
\tablehead{
\colhead{Name} &
\colhead{$z_{\rm [CII]}$} &
\colhead{$z_{\rm [OIII]}$\tablenotemark{a}} &
\colhead{$M_{\rm BH, H\beta}$} &
\colhead{FWHM$_{\rm H\beta}$} &
\colhead{$L_{\rm 5100}$} &
\colhead{$\lambda_{\rm Edd}$} &
\colhead{$M_{\rm BH, MgII}$} &
\colhead{Ref\_\mgii\tablenotemark{b}} 
\\
\nocolhead{} & \nocolhead{} & \nocolhead{} & \colhead{($10^{9}\,M_{\odot}$)}  & \colhead{(km s$^{-1}$)} &  \colhead{($10^{45} $\rm erg s$^{-1}$)} & \nocolhead{} & \colhead{($10^{9}\,M_{\odot}$)} & \nocolhead{}
}
\startdata
J010953.13--304726.30 & 6.7904$\pm$0.0003 & -- & 0.60$\pm$0.05 & 3033.0$\pm$117.0 & 6.40$\pm$0.08 & 0.76$\pm$0.06 & 1.11$^{+0.4}_{-0.36}$ & Farina+2022 \\
J021847.04+000715.20 & 6.7700$\pm$0.0013 & 6.7668$\pm$0.0006 & 0.61$\pm$0.05 & 3030.0$\pm$117.0 & 6.71$\pm$0.16 & 0.78$\pm$0.07 & 0.61$\pm$0.07 & Yang+2021 \\
J022426.54--471129.40 & 6.5222$\pm$0.0001 & 6.5186$\pm$0.0004 & 2.15$\pm$0.09 & 3936.0$\pm$78.0 & 29.19$\pm$0.09 & 0.97$\pm$0.04 & 1.30$\pm$0.18 & Wang+2021 \\
J022601.87+030259.28 & 6.5405$\pm$0.0001 & -- & 1.05$\pm$0.09 & 3131.0$\pm$129.0 & 17.25$\pm$0.09 & 1.17$\pm$0.09 & 2.20$\pm$0.39 & Wang+2021 \\
J024401.02--500853.70 & 6.7306$\pm$0.0002 & 6.720$\pm$0.003 & 0.88$\pm$0.10 & 2969.0$\pm$158.0 & 15.08$\pm$0.13 & 1.22$\pm$0.12 & 1.15$\pm$0.39 & Reed+2019 \\
J030516.92--315056.00 & 6.6139$\pm$0.0002 & 6.615$\pm$0.001 & 0.62$\pm$0.14 & 3020.0$\pm$314.0 & 6.95$\pm$0.06 & 0.80$\pm$0.14 & 0.70$\pm$0.38 & Wang+2021 \\
J200241.59--301321.69 & 6.6876$\pm$0.0004 & -- & 0.84$\pm$0.04 & 2981.0$\pm$79.0 & 13.65$\pm$0.10 & 1.15$\pm$0.06 & 1.62$\pm$0.27 & Yang+2021 \\
J223255.15+293032.04 & 6.666$\pm$0.004 & 6.669$\pm$0.0001 & 1.93$\pm$0.13 & 6044.0$\pm$201.0 & 4.24$\pm$0.10 & 0.16$\pm$0.01 & 3.06$\pm$0.36 & Yang+2021 \\
\enddata
\tablenotetext{a}{The [\oiii] redshift is estimated based on the line centered of the `core' component of [\oiii] 5008 line. Quasars J0109--3047, J0226+0302, and J2002--3013 do not have narow [\oiii] components, and thus we do not estimate their [\oiii] redshift. }
\tablenotetext{b}{The references of the \mgii-based BH masses adopted in this work.}
 \label{tab:tab1}
\end{deluxetable*}

We measure the single-epoch virial BH masses of these quasars based on their continuum luminosity at rest-frame 5100 \AA\ and the H$\beta$ line width, obtained from the spectral fitting. We adopt the BH mass estimator in \cite{vp06}, 
\begin{equation}
    \frac{M_{\rm BH, H\beta}}{M_\sun} = 10^{6.91} \left(\frac{\rm FWHM(H\beta)}{1000\,\rm km\,s^{-1}}\right)^{2}\left(\frac{\lambda L_{\lambda}(5100 \rm \AA)}{10^{44}\,\rm erg\,s^{-1}}\right)^{0.5}.
\end{equation} 
The 5100 \AA\ continuum luminosity is derived from the best-fit power law continuum directly. The systematic scatter of this BH mass estimator relative to RM mass is $\sim$ 0.4 -- 0.5 dex, which is not included when we report the errors of BH mass measurements. The uncertainties of all measurements are derived from the MC approach, as described above. 
We then compute the Eddington ratio ($\lambda_{\rm Edd}$ = $L_{\rm bol}/L_{\rm Edd}$) for each quasar. The bolometric luminosity is estimated by multiplying the 5100 \AA\ luminosity by 9.26 \citep{richards06}.
Table 1 summarizes the redshift, H$\beta$ line width, continuum luminosity, H$\beta$ BH mass, Eddington ratio, and \mgii\ BH mass of each of the eight quasars. 

\begin{figure}
\centering 
\epsscale{1.22}
\plotone{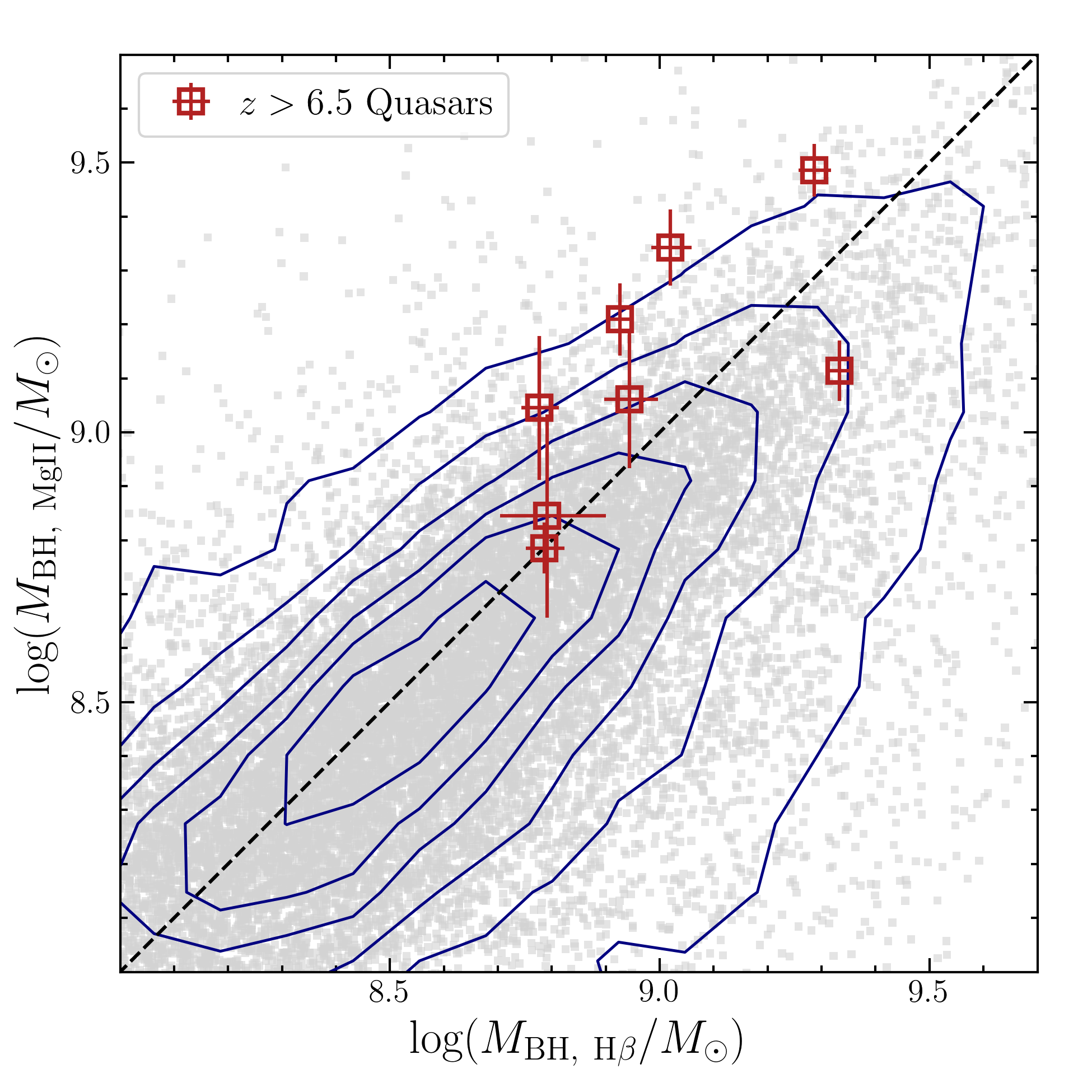} 
\caption{The BH masses of eight ASPIRE quasars (red squares), measured from the H$\beta$ lines in JWST/NIRCam WFSS spectra and the \mgii\ lines in their NIR spectra. These are compared with the measurements of low-$z$ quasars from the SDSS DR7 sample \citep{shen11} using the same \mgii\ and H$\beta$ BH mass estimators (grey small squares and blue contours). The differences between $M_{\rm BH, H\beta}$ and $M_{\rm BH, MgII}$ of these high-redshift quasars are comparable to the scatter shown in the low-redshift quasar sample. The mean value of $M_{\rm BH, H\beta}$ - $M_{\rm BH, MgII}$ is $-0.13$ dex, while the sample size is too small to identify a systematic offset.}
\label{fig:mbh}
\end{figure}

The H$\beta$-based measurements show that these quasars have BH masses in the range of (0.6 -- 2.1)$\times$$10^{9}\,M_{\odot}$. 
These quasars all have been published with measurements of BH mass based on the \mgii\ emission lines from their NIR spectroscopic data \citep[e.g.,][]{mazzucchelli17, reed19,schindler20,wang21,yang21,farina22}. We adopt the most recent measurements of these eight quasars from \cite{reed19}, \cite{wang21}, \cite{yang21}, and \cite{farina22}. All of these \mgii-based BH masses are computed using the BH mass estimator from \cite{vo09}. Figure \ref{fig:mbh} shows the H$\beta$-based BH masses compared to their \mgii-based BH masses. The BH masses derived from the H$\beta$ lines and 5100 \AA\ luminosity are consistent with the \mgii-based measurements within their systematic scatters (i.e., $\sim$ 0.4 -- 0.5 dex). Recently, a new measurement of BH mass of quasar J0100+2802 at $z=6.3$ using H$\beta$ line from JWST observations also shows agreement with its \mgii-based BH mass within systematic uncertainties \citep{eilers22}. 

In our quasar sample, the differences between these two BH mass estimates span from $-$0.32 to 0.22 dex, comparable to the scatter shown in the SDSS low-redshift quasar sample \citep[a standard deviation of $\sim$ 0.38 dex,][]{shen11}. Among these eight quasars, there is a trend that the H$\beta$-based BH masses are slightly smaller, with an average $M_{\rm BH, H\beta}$ -- $M_{\rm BH, MgII}$ of $-0.13$ dex. However, the current sample is too small to conclude if there is a systematic offset. The H$\beta$-based BH masses and the comparison between H$\beta$ and \mgii\ measurements confirm the existence of billion solar-mass BHs in the reionization epoch.
These SMBHs require $\gtrsim 10^{2-3}~M_{\odot}$ seed BHs at $z=30$ assuming Eddington accretion across the entire time since BH seeding and a radiative efficiency of 0.1 \citep[e.g.,][]{yang21, farina22}. This thus raises the questions of how to maintain highly accreting BHs for a long term and how to form massive seed BHs in the early Universe \citep[e.g.,][]{volonteri12,davies19,inayoshi20}.

These measurements are derived from the spectral fits of the NIR spectra and JWST spectra separately. Thus, for both data sets, the limited continuum windows might lead to additional uncertainties of the spectral decomposition and BH mass measurements. In the next step, we will utilize the entire ASPIRE sample to perform a detailed comparison between the H$\beta$ and \mgii\ BH masses when observations of the full sample are complete. With the final sample, we will combine the NIR and JWST spectra as well as the JWST F115W, F200W, and F356W photometry, and will apply joint spectral analysis. The combined dataset will provide a wider continuum window and allow a better separation of the \feii\ emission from the featureless continuum. 

\subsection{Optical Spectral Properties and [\oiii] Outflows}
\begin{figure}
\centering 
\epsscale{1.2}
\plotone{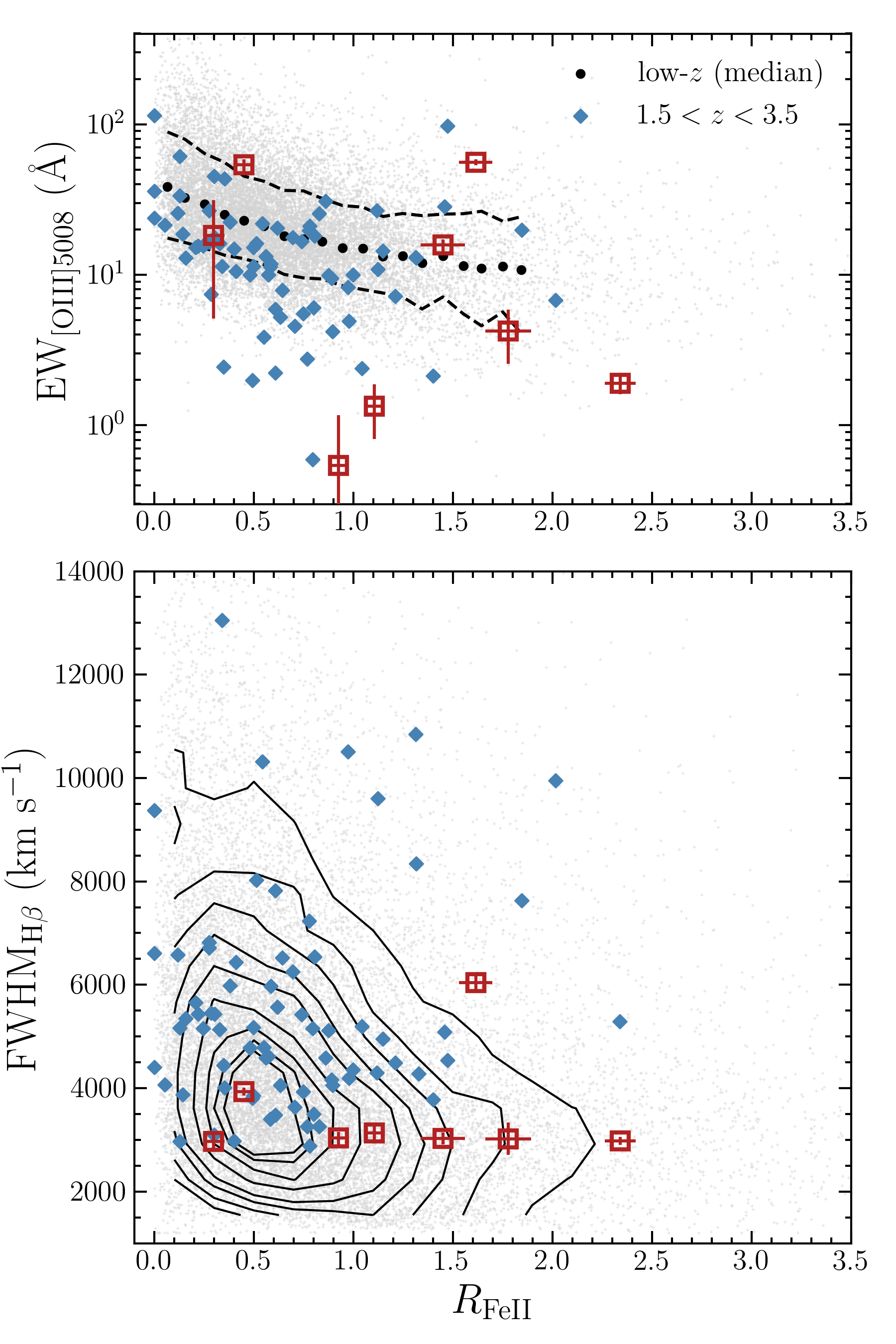} 
\caption{EV1 relations using our eight high-redshift quasars (red squares), compared with luminous quasars at intermediate redshift \citep[][open diamonds]{shen16} and SDSS low-z quasars \citep[][grey dots]{shen11}. The black filled circles and the dashed lines denote the median, 16th and 84th percentiles of the SDSS low-z quasars at each redshift bin. These reionization-era quasars generally follow the trends observed in lower-redshift quasars.}
\label{fig:ev1}
\end{figure} 

The most important spectral quantities covered by the WFSS spectra are H$\beta$, [\oiii], and \feii\ emission. In general, all eight quasars have strong and broad H$\beta$ lines, with FWHM ranging from 3000 to 6000 km s$^{-1}$. Optical \feii\ emission is clearly seen in all eight quasars. The observed [\oiii] doublets in these quasars have relatively diverse properties. Three of these quasars, J0109--3047, J0226+0302, and J2002--3013, show similar profiles in their [\oiii] line regions, with very weak broad-only [\oiii] doublets but no narrow [\oiii] lines (i.e., the `core' component, $< 1200$ km s$^{-1}$). 
The other five quasars all have both narrow and broad [\oiii] components. 
Quasar J2232+2930 has the strongest narrow [\oiii] lines among these quasars. It is the only object observed with stronger narrow [\oiii] lines relative to its broad components.

Using the WFSS data, we are able to visit the Eigenvector 1 (EV1) relations \citep{BG92} in the most distant quasars. The EV1 relations describe the relationships between emission-line quantities and the strength of the optical \feii\ emission, and has been widely investigated in the past three decades with quasar samples at $z < 4$ \citep[e.g.,][]{BG92,boroson02,marziani01,sh14}. The anti-correlation between the strength of [\oiii] and \feii\ emission is one of the most important EV1 relations observed. The trend of \feii\ strength, as the H$\beta$ FWHM decreases, is also a probe of the physical drivers of the EV1 relations. We study the correlation of the [\oiii] 5008 EW and FWHM$_{\rm H\beta}$ with the $R_{\rm FeII}$ \citep{shen16}, and compare them with the observations of the luminous intermediate-redshift ($1.5 < z < 3.5$) sample in \cite{shen16} and the low-redshift ($z< 1$) quasars from the SDSS DR7 \citep{shen11}.
The EW$_{\rm [OIII]5008}$ used here is the EW of the entire [\oiii] $\lambda$5008 line, in order to compare our results with the measurements from \cite{shen16}. FWHM$_{\rm H\beta}$ is the line width of the broad H$\beta$. 
$R_{\rm FeII}$ is the ratio of EW$_{\rm FeII}$ and EW$_{\rm H\beta}$, and the \feii\ EW is measured between 4434 and 4684 \AA. 

As shown in Fig. \ref{fig:ev1}, overall, the optical spectral quantities in these high-redshift quasars follow the relations defined in lower-redshift quasars. Our high-redshift quasars show slightly lower [\oiii] EW than the mean values of SDSS low-redshift quasars and are more similar to the luminous $1.5 < z < 3.5$ quasar sample, which is due to their high luminosity and the Baldwin effect \citep{baldwin77}. There is no significant outlier among our high-redshift quasars. The three lowest [\oiii] EWs are from the quasars with weak and broad-only [\oiii] emission. In the FWHM$_{\rm H\beta}$ vs. $R_{\rm FeII}$ plane, our objects distribute in a narrow FWHM$_{\rm H\beta}$ range with lower FWHM$_{\rm H\beta}$ than the intermediate-redshift sample. This is consistent with the high Eddington ratios of most (7 of 8) of our high-$z$ quasars. 
The general consistency between our sample and low-$z$ distributions suggests that the EV1 relations might exist in quasars as early as the reionization epoch, and our full ASPIRE sample will enable more detailed comparisons with low-$z$ samples. 

\begin{figure}
\centering 
\epsscale{1.2}
\plotone{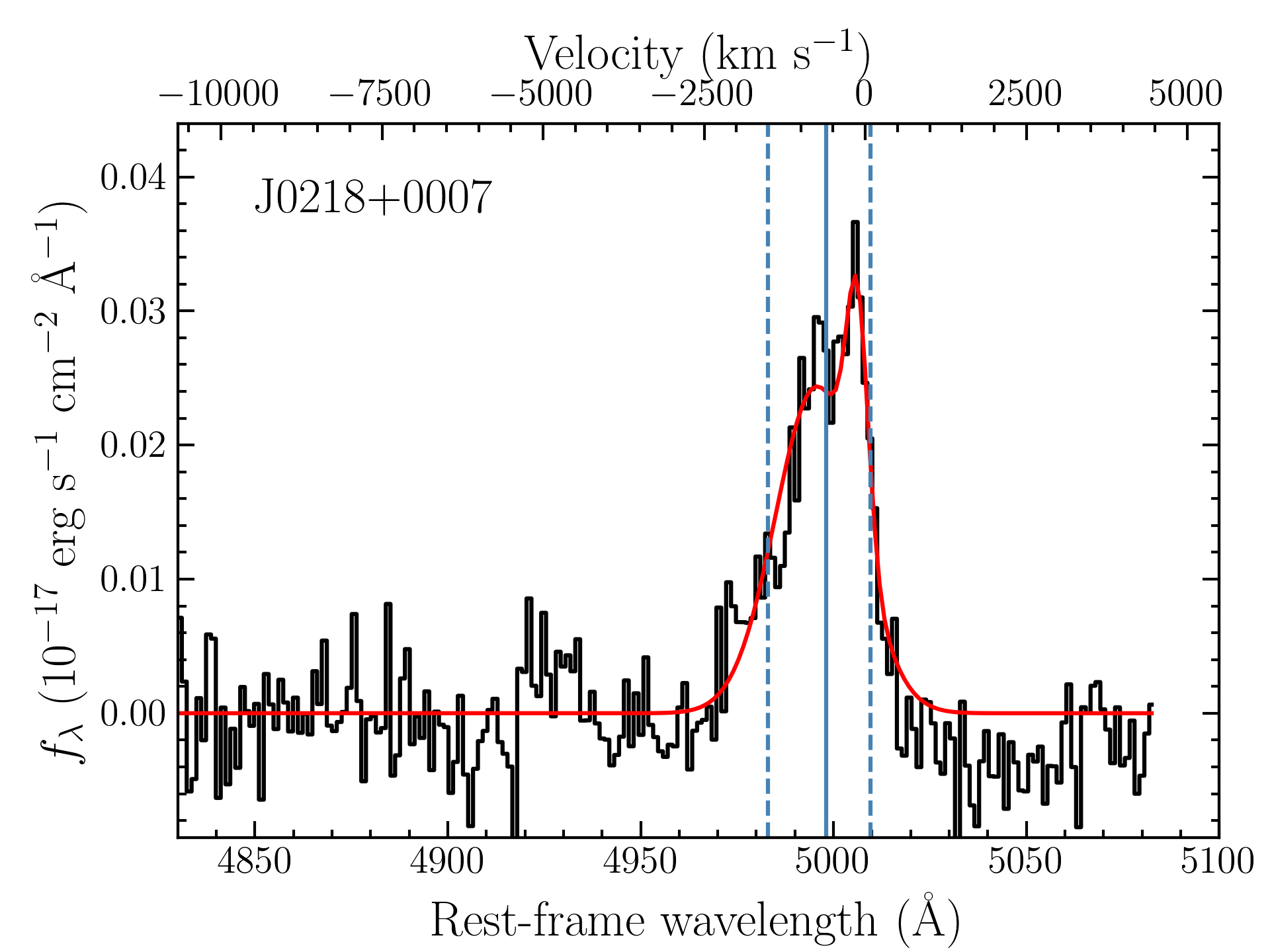} 
\plotone{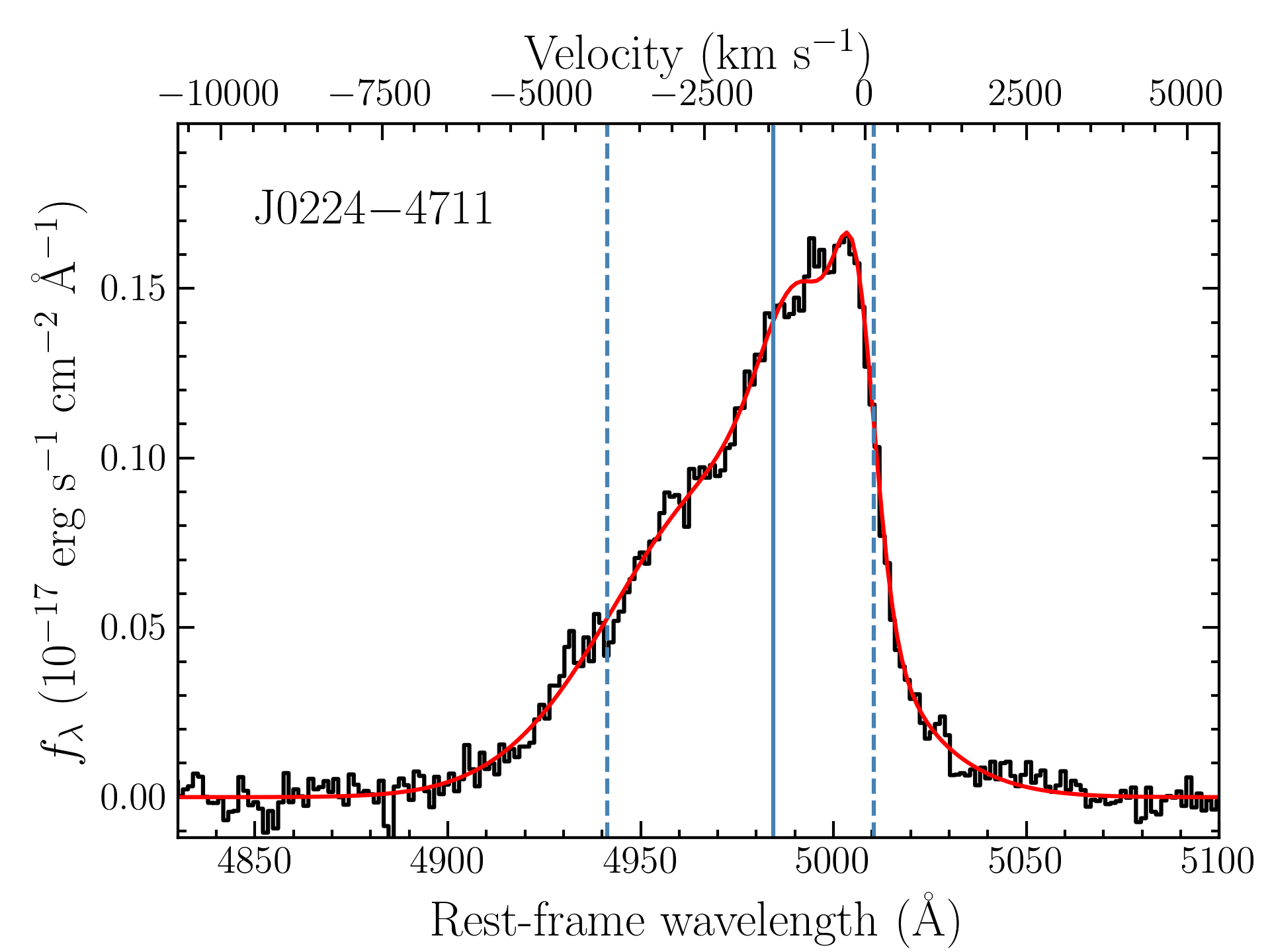} 
\caption{The [\oiii] 5008 lines with strong broad and blueshifted components have been observed in two quasars, J0218+0007 (top) and J0224--4711 (bottom). The blue solid lines denote the median velocity ($v_{\rm 50}$), and the dashed lines mark the part containing 80 percent of the line power ($w_{80}$). J0218+0007 has a $v_{\rm 50}$ of $\sim $ --610 km s$^{-1}$ and a $w_{80}$ of 1590 km s$^{-1}$, Quasar J0224--4711 has a more remarkable [\oiii], with a $v_{\rm 50}$ of $\sim $ --1430 km s$^{-1}$ and a $w_{80}$ of 4140 km s$^{-1}$. These blueshifted broad lines, in particular in J0224--4711, are comparable to the most extreme cases observed at lower redshift \citep[e.g.,][]{zakamska16,bischetti17}.}
\label{fig:oiii}
\end{figure}

Broad, skewed, and blue/redshifted [\oiii] profiles have been discovered in quasars at $z \lesssim 3.5$ and used to probe kpc-scale outflows in quasar host galaxies \citep[e.g.,][]{cano-diaz12, carniani15, zakamska16, bischetti17}.
At $z > 6$, outflows powered by the central AGN are a common prediction of a broad range of cosmological simulations \citep[e.g.,][]{dubois13, costa14, costa22, lupi22}. However, observational evidence of large-scale outflows in quasar hosts remains anecdotal at this redshift. 
The JWST dataset provides us with the first chance to search for [\oiii] outflows in reionization-era quasars. Two quasars, J0218+0007 and J0224--4711, have prominently strong broad and blueshifted [\oiii] components, with a velocity shift of $-630$ km s$^{-1}$ and $-1690$ km s$^{-1}$ relative to the narrow [\oiii], respectively. Figure \ref{fig:oiii} shows their zoomed-in [\oiii] $\lambda$5008 lines, with other components subtracted out according to the best fit models. 
We compute the median velocity, $v_{\rm 50}$, the velocity at the 50th percentile of the line flux; $v_{\rm 10}$ and $v_{\rm 90}$, the velocities at the 10th and 90th percentiles; $w_{80}$ (= $v_{\rm 90}$ - $v_{\rm 10}$), containing 80 percent of the total line power, which is close to the commonly used FWHM ($w_{80}$ = 1.088 FWHM) for a Gaussian profile. The velocities are calculated relative to their [\cii] redshift. 
J0218+0007 has a median velocity ($v_{\rm 50}$) of $\sim -610$ km s$^{-1}$, with a $v_{\rm 90}$ of $-$1510 km s$^{-1}$ and a $w_{80}$ of 1590 km s$^{-1}$. 
J0224--4711 has a median velocity ($v_{\rm 50}$) of $\sim -1430$ km s$^{-1}$, a $v_{\rm 90}$ of $-$4010 km s$^{-1}$, and a $w_{80}$ of $\sim 4140$ km s$^{-1}$.

Their broad [\oiii] widths and high velocity shifts make them stand out among the quasar population. Particularly, J0224--4711 has one of the most extreme broad and blueshifted [\oiii] lines observed to date, even compared to the observations of lower-redshift quasars. For example, the $1.5 < z < 3.5$ luminous quasars in \cite{shen16} have a median [\oiii] FWHM of 900 km s$^{-1}$, with the broadest one to 2100 km s$^{-1}$. 
\cite{zakamska16} observed extreme outflows in red luminous quasars at $z \sim 2-3$ and discovered broad [\oiii] with $w_{80}$ ranging between 3600 and 5500 km s$^{-1}$, strongly blueshifted up to 1500 km s$^{-1}$. The more recent study of hyper-luminous quasars at $z \sim 2-4$ revealed a population of [\oiii] lines with FWHM of 1200--2200 km s$^{-1}$ \citep{bischetti17}. Large velocity widths and high velocities of blueshifted [\oiii] suggest powerful ionized outflows, while the nature of high-velocity large-scale gas in quasars is still debated. 
The most blueshifted [\oiii] lines observed in the extremely red $z=2-3$ quasars are thought to be caught during a `blow-out' phase of quasar evolution and associated with fast ionized outflows \citep[e.g.,][]{zakamska16, perrotta19, vayner21}. 
So far, there is no evidence of J0218+0007 and J0224--4711 being extremely red quasars. 
A detailed multi-wavelength analysis will be conducted in the future to better understand the physical properties of their fast outflows. 
In addition, a joint analysis of the 1D spectra and 2D images will be performed in subsequent works to examine the host galaxies and outflows. In the next step, JWST NIRSpec/IFU follow-up observations will be carried out to reveal the nature of their outflows. The entire ASPIRE sample will allow us to systematically study [\oiii] outflows in the reionization-era quasars, in synergy with multi-wavelength imaging and spectroscopic data. 

\section{Summary}
In this paper, we present the results from an early sample of ASPIRE JWST/WFSS spectral dataset consisting of eight quasars at $z>6.5$, observed in 2022 August and September. With these high quality WFSS spectra, we are able to study the rest-frame optical emission in reionization-era quasars for the first time. We measure the virial BH masses using the H$\beta$ lines and compare them with the \mgii-based measurements. We find that in general the H$\beta$ BH masses are consistent with the \mgii\ BH masses within the systematic uncertainties of the BH mass estimators. The discrepancy is within the scatter between these two measurements as gauged from lower-redshift quasar samples. Therefore, using a more reliable tracer, the H$\beta$ line, we confirm the existence of billion solar-mass BHs in the reionization epoch. 

We also test the optical spectral properties of these reionization-era quasars in the EV1 plane defined by $z< 4$ quasars. The results suggest that the EV1 relations exist in these distant quasars and our quasars are distributed similarly to the luminous quasars at intermediate redshift. We discover two quasars with broad and strongly blueshifted [\oiii] lines. One of them has a prominent broad [\oiii] profile ($w_{80}$ = 4140 km s$^{-1}$) with a high velocity shift ($v_{\rm 50}$ = $-$1430 km s$^{-1}$), which is among the most extreme cases, even compared to observations in lower-redshift quasars. 

With the ASPIRE program, we will construct a sample of 25 quasars with not only JWST WFSS spectra and NIRCam images but also the data from multi-wavelength observations, spanning X-ray to submillimeter and radio. The final sample will be used for detailed  investigations of their BH physics, quasar feedback, and host galaxy properties.

\acknowledgments
J. Yang and X. Fan acknowledge support from US NSF grants AST 19-08284.
F. Wang acknowledges support by NASA through the NASA Hubble Fellowship grant \#HST-HF2-51448.001-A awarded by the Space Telescope Science Institute, which is operated by the Association of Universities for Research in Astronomy, Incorporated, under NASA contract NAS5-26555.
Research at UC Irvine was supported by NSF grant AST-1907290.
F.S. and E.E. acknowledge funding from JWST/NIRCam contract to the University of Arizona, NAS5-02105. 
JTS acknowledges funding from the European Research Council (ERC) Advanced Grant program under the European Union's Horizon 2020 research and innovation programme (Grant agreement No. 885301).
LB acknowledges support from NSF award AST-1909933 and NASA award \#80NSSC22K0808.
SEIB acknowledges funding from the European Research Council (ERC) under the European Union's Horizon 2020 research and innovation programme (grant agreement no. 740246 ``Cosmic Gas''.
M.H. acknowledges support from the Zentrum f\"ur Astronomie der Universit\"at Heidelberg under the Gliese Fellowship.
ZH acknowledges support from NSF grant AST-2006176.
GK is partly supported by the Department of Atomic Energy (Government of India) research project with Project Identification Number RTI~4002, and by the Max Planck Society through a Max Planck Partner Group.
AL acknowledges funding from MIUR under the grant PRIN 2017-MB8AEZ.
S.R.R. acknowledges financial support from the International Max Planck Research School for Astronomy and Cosmic Physics at the University of Heidelberg (IMPRS-HD).
BT acknowledges support from the European Research Council (ERC) under the European Union's Horizon 2020 research and innovation program (grant agreement 950533) and from the Israel Science Foundation (grant 1849/19).
MT acknowledges support from the NWO grant 0.16.VIDI.189.162 (``ODIN'').
MV gratefully acknowledges support from the Independent Research Fund Denmark via grant number DFF 8021-00130. 

This work is based on observations made with the NASA/ESA/CSA James Webb Space Telescope. 
These observations are associated with program \#2078. Support for program \#2078 was provided by NASA through a grant from the Space Telescope Science Institute, which is operated by the Association of Universities for Research in Astronomy, Inc., under NASA contract NAS 5-03127. All of the data presented in this paper were obtained from the Mikulski Archive for Space Telescopes (MAST) at the Space Telescope Science Institute. The specific observations analyzed can be accessed via \dataset[DOI]{https://doi.org/10.17909/vt74-kd84}.

%

\vspace{5mm}
\facilities{JWST(NIRCam)}


\software{astropy \citep{Astropy},  
Matplotlib \citep{Matplotlib},
Numpy \citep{Numpy},
Photutils \citep{photutils},
Scipy \citep{Scipy}
}








\end{document}